\documentclass[twoside,11pt]{article}

\usepackage[preprint]{jmlr2e}
\usepackage{mystyle}

\ShortHeadings{TensorLy-Quantum: Quantum Machine Learning with Tensor Methods}{Patti, Kossaifi, Yelin and Anandkumar}
\firstpageno{1}

\begin{document}

\title{TensorLy-Quantum: Quantum Machine Learning with Tensor Methods}

\author{\name Taylor L. Patti\(^{1,2}\) \email tpatti@nvidia.com\\
\name Jean Kossaifi\(^{1}\) \email jkossaifi@nvidia.com\\
\name Susanne F. Yelin\(^{2}\) \email syelin@g.harvard.edu\\
\name Anima Anandkumar\(^{1,3}\) \email anima@caltech.edu\\
\addr \(^{1}\)NVIDIA\\
\addr \(^{2}\)Harvard University\\
\addr \(^{3}\) California Institute of Technology
}

\editor{-}

\maketitle

\begin{abstract}
Simulation is essential for developing quantum hardware and algorithms. However, simulating quantum circuits on classical hardware is challenging due to the exponential scaling of quantum state space. While factorized tensors can greatly reduce this overhead, tensor network-based simulators are relatively few and often lack crucial functionalities. To address this deficiency, we created TensorLy-Quantum, a Python library for quantum circuit simulation that adopts the PyTorch API. Our library leverages the optimized tensor methods of the existing TensorLy ecosystem to represent, simulate, and manipulate large-scale quantum circuits. Through compact tensor representations and efficient operations, TensorLy-Quantum can scale to hundreds of qubits on a single GPU and thousands of qubits on multiple GPUs. TensorLy-Quantum is open-source and accessible at https://github.com/tensorly/quantum 
\end{abstract}


\section{Introduction}

The size and intricacies of quantum systems have long motivated quantum computing (QC) research~(\cite{montanaro:16}). Recently, quantum algorithms have incorporated variational techniques~(\cite{peruzzo:14,farhi:14}), creating the subfield of quantum machine learning (QML)~\cite{mehta:19}, which includes applications like optimization and quantum chemistry. Much like its classical counterpart, QML leverages gradient-based methods to seek optimal solutions, but with the added benefit of an exponentially large solution kernel~(\cite{schuld:21}). For instance, the most common form of QML uses quantum expectation values to define objective functions and classical gradient descent to optimize quantum networks. Both QC and QML ultimately seek to demonstrate a tangible improvement over classical computing for useful tasks, a concept known as quantum advantage.

As QC and QML are still nascent fields, simulation remains essential to both the characterization of quantum systems at scale~(\cite{carleo:19}), as well as the design heuristic algorithms~(\cite{cerezo:21}). However, simulating quantum algorithms on informative scales is a challenging task, as quantum state space overwhelms classical resources with even relatively few quantum bits (qubits). This limitation is partially overcome with networks of factorized tensors~(\cite{bridgeman:17}), which are typically more efficient than the traditional vector representation of quantum states, providing up to an exponential reduction in overhead with respect to the number of qubits $n$. In tensor network-based simulation, both quantum states, gates, and operators are expressed in a factorized form, which compress the quantum circuit by including only relevant ranks. Operations can be efficiently done directly in the factorized form without having to reconstruct any dense tensor. The individual factors are contracted, or summed over at corresponding indices, along an optimized contraction path. These paths are typically chosen to minimize either memory or runtime~(\cite{smith:18}). Quantum tensor contractions consist of highly parallelizable and arithmetically intensive operations, making them prime candidates for GPU acceleration~(\cite{cuTensor}). A similar approach has proven useful in deep learning, for instance by expressing the weights of linear~(\cite{novikov:15}) and convolutional~(\cite{kossaifi2020factorized}) layers of deep nets in factorized form. However, this approach remains under-explored in quantum simulation. 


\begin{figure}[!b]
\centering
\includegraphics[width=0.93\textwidth]{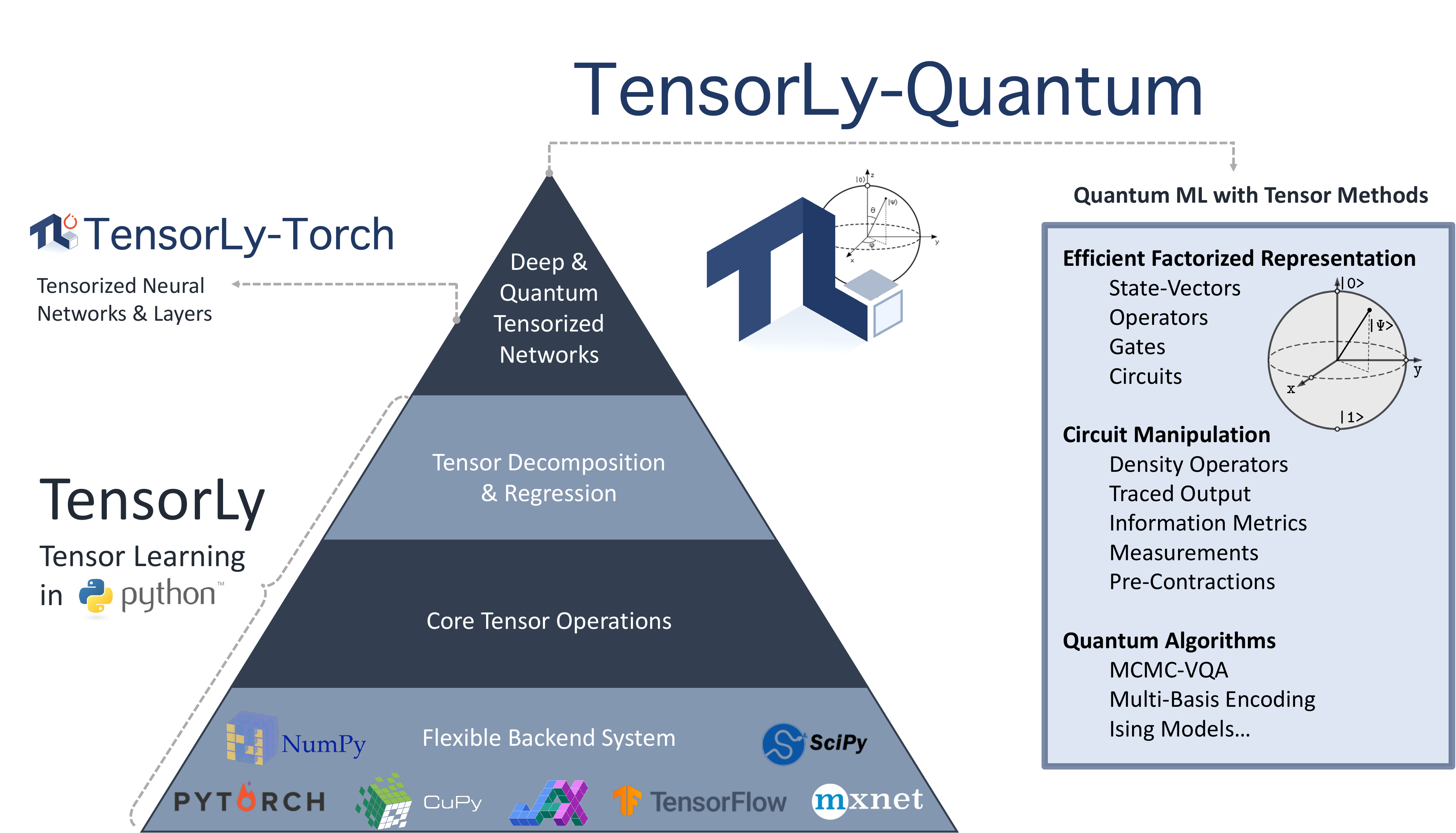}
\caption{TensorLy-Quantum sits atop the hierarchy of TensorLy libraries, inheriting its extensive tensor methods and providing full Autograd support for the features (circuits, operations, density operators, and algorithms) required by quantum simulation.}
\label{fig:diagram}
\end{figure}

While a variety of quantum simulators exist~(\cite{oakridge,qiskit,rigetti}), relatively few utilize tensor methods. Among those that do, integration with essential machine learning tools, such as automatic differentiation (i.e., PyTorch Autograd), is rare~(\cite{gray:18}). Moreover, advanced functionalities, like tensor regression, while used in machine learning~(\cite{panagakis2021tensor}), are so far unused in quantum simulation.

To address such deficits, we have created TensorLy-Quantum, a PyTorch API for efficient tensor-based simulation of QC and QML protocols. It is a member of the TensorLy family of libraries and makes use of its extensive features and optimized implementations~(\cite{kossaifi:19}). As a result, it is the first quantum library with direct support for tensor decomposition, regression, and algebra, which have proven fruitful in a myriad of fields and stand to enrich QC and QML research. Uniquely, TensorLy-Quantum provides built-in support for Multi-Basis Encoding for MaxCut problems~(\cite{patti:21}) and was used to develop Markov Chain Monte Carlo-based QML~(\cite{pattiMCMC:21}). Through these features, TensorLy-Quantum aims to provide extensive tensor-based quantum simulation capabilities, providing a simple and flexible API for quantum algorithms research. Moreover, its optimized functionality and lightweight API would make it an efficient QML backend to supplement other quantum simulators.

\section{TensorLy-Quantum}

TensorLy-Quantum offers functionalities essential to quantum computation (Fig.~\ref{fig:diagram}). The library provides a high-level interface for quantum circuit simulation, with an API that follows the PyTorch Module structure and offers end-to-end support for automatic differentiation via PyTorch Autograd. 
Users can seamlessly design quantum circuits by combining either pre-defined or customizable quantum gates, operators, and states. 
The library also provides extensive circuit operations, ranging from pre-contraction techniques that simplify contraction path search, to dynamic partial traces that compactly evaluate quantum circuit outputs. In addition to specializing in factorized representations, such as Matrix-Product State (known as tensor-train in the Machine-Learning literature~\cite{oseledets2011tensor}), TensorLy-Quantum also supports efficient analysis on quantum density operators, both pure and reduced, including partial traces and information metrics. TensorLy-Quantum is designed to bridge the gap between practitioners of quantum and classical machine learning, providing an intuitive and Pythonic interface that is supplemented with extensive documentation and 97\% unit-test coverage.

TensorLy-Quantum leverages the deep tensorized network capabilities of TensorLy and TensorLy-Torch, using these to mitigate the significant computational overhead posed by quantum state space. It is suitable for both CPUs and GPUs, and that it acquires the GPU acceleration of these parent libraries. Likewise, while TensorLy-Quantum is PyTorch-based, TensorLy's flexible backend structure enables dynamic transitions between many of the most utilized Python libraries for machine learning and numerical analysis. Due to its strategic location atop the TensorLy ecosystem (Fig.~\ref{fig:diagram}), TensorLy-Quantum is exceptionally positioned to accelerate and innovate quantum simulation. In what follows, we illustrate the scalability and speed of our library, particularly on GPU.

\section{Performance}

Due to its efficiency and scalability, TensorLy-Quantum holds world records for the number of qubits simulated in a successful quantum optimization algorithm. These records include both the largest single-qubit implementation of MaxCut~(\cite{patti:21}) and a multi-GPU scaleup that used cuQuantum as a backend for tensor network contraction~(\cite{NVIDIA:21, cuQuantum}).

\begin{figure}
\centering
\captionsetup[subfigure]{labelformat=empty}
\quad
\subfloat[\centering]{{\includegraphics[height=0.4\textwidth]{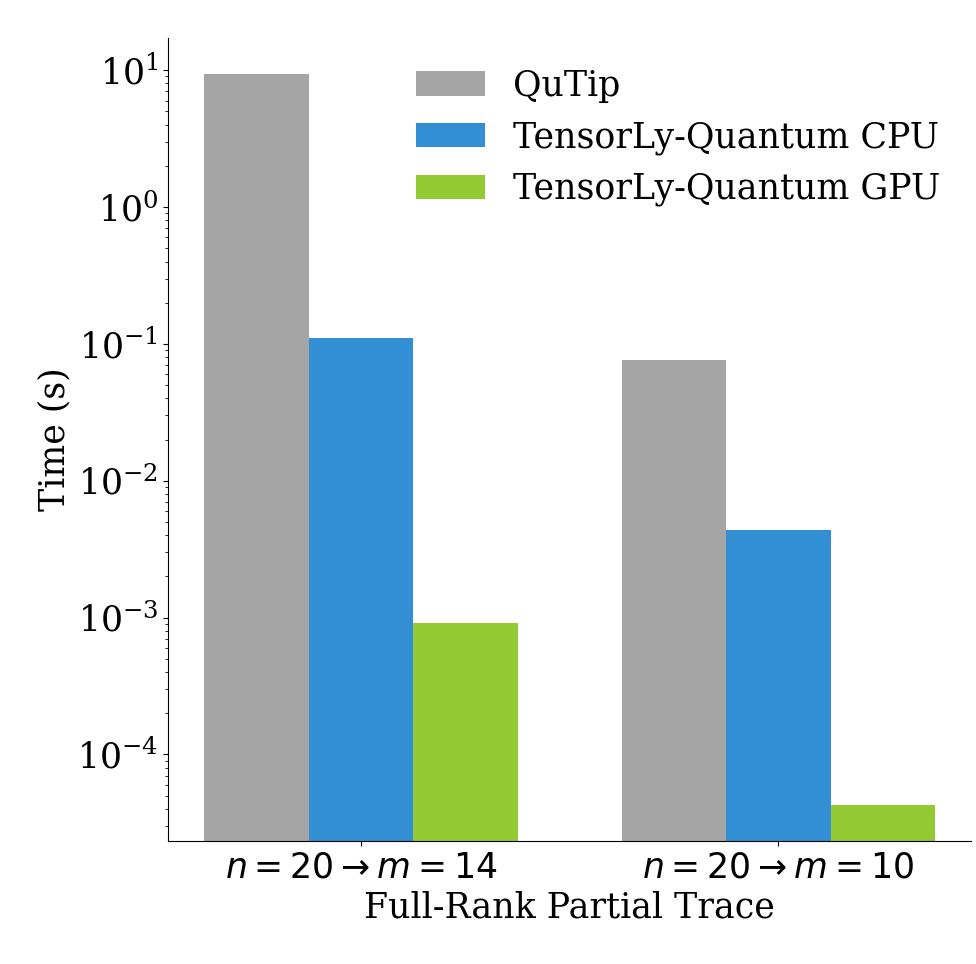} }}
\hfill
\subfloat[\centering]{{\includegraphics[height=0.4\textwidth]{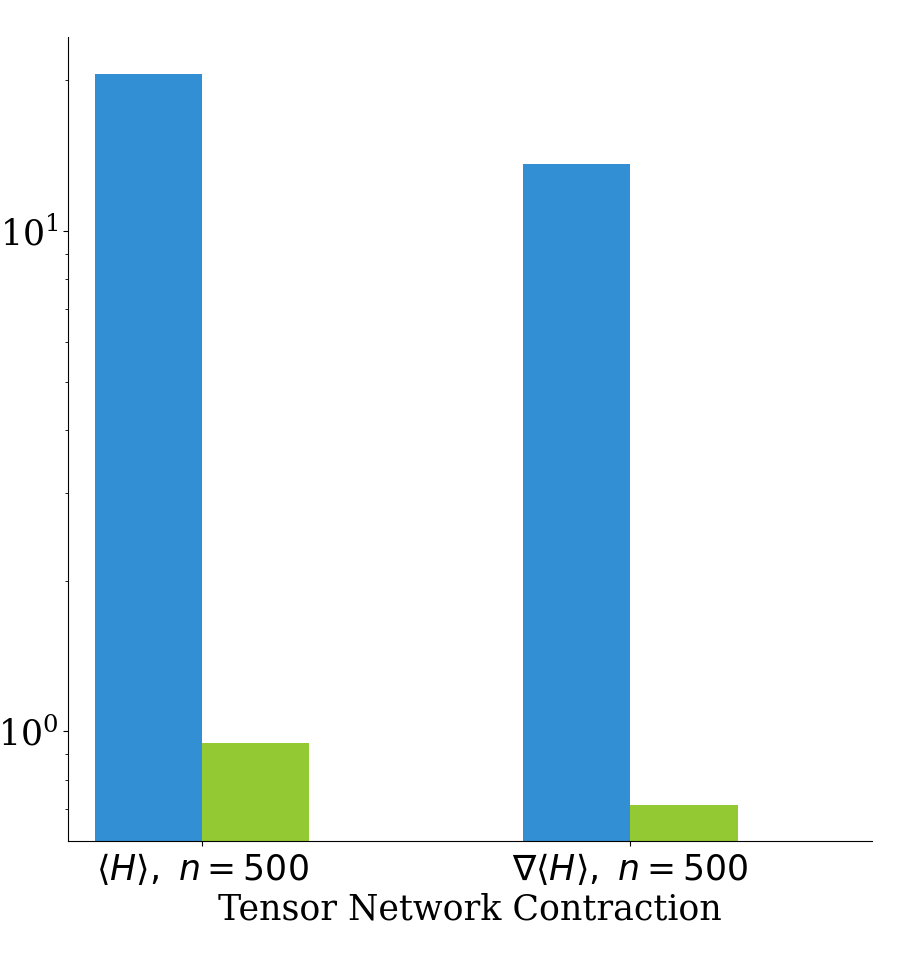} }}
\quad
\vspace{-20pt}
\caption{Logarithmic-scale runtime comparison for (left) full-rank partial trace from $n$ to $m$ qubits and (right) contraction operations on large-scale networks of factorized tensors.}
\label{fig:1}
\end{figure}

We highlight this performance with numerical experiments, showcasing both operations on the full-rank density operator (a reshaped, matrix-like quantum state representation) and tensor contraction functionalities of TensorLy-Quantum. In the density operator experiments, TensorLy-Quantum outperforms the leading software, QuTip~(\cite{johansson:12}), by two orders of magnitude on CPU for partial traces of $n=20$ to $m=14,10$ qubit systems and four orders of magnitude on GPU. TensorLy-Quantum can also complete full-rank calculations on larger quantum systems than its predecessors through the use of compact tensor algebra.

Experiments on networks of factorized tensors include a forward pass of the expectation value $\langle H \rangle$, and the full gradient calculation $\nabla \langle H \rangle$, where $H$ is a transverse-field Ising model Hamiltonian of $n=500$ qubits. The circuit ansatze contain $5000$ gates and the gradient calculation constitutes a full backpropagation with PyTorch Autograd. GPU acceleration provides a $20$x speedup over the CPU implementation. Moreover, we emphasize that without factorized tensor methods, simulations of this size are impossible, as matrix-encoded Hamiltonians require $\sim 10^6$ GB of memory for systems of merely $n \sim 25$ qubits. Contraction was accomplished with the Opt-Einsum library~(\cite{smith:18}) and all GPU experiments were done on an NVIDIA A100 GPU.

\section{Conclusions}
TensorLy-Quantum is an open-source library designed to streamline the workflow of QC and QML researchers. It provides highly optimized operations for large-scale and compute heavy quantum circuit simulations, buttressed by extensive documentation and high-coverage unit-testing. Its API seamlessly integrates with PyTorch and provides an interface that is amenable to diverse scientific backgrounds, accessibly packaging state-of-the-art tensor network operations alongside quantum protocols. Moreover, its lightweight API and optimized functions would make it an excellent backend for existing quantum simulators. As TensorLy-Quantum uses the TensorLy libraries as a backend, it offers direct access to tools unavailable in other quantum APIs, like tensor regression and decomposition, as well as convenient conversion to the native data structures of numerous backends. In future releases, we will expand both the efficiency and scope of TensorLy-Quantum, adding features such as causality-simplified contraction, quantum state compression, and novel quantum protocols, as well as support for more factorized representations, using TensorLy's existing tensor decomposition.








\vskip 0.2in
\bibliography{sample}

\end{document}